# A RAPID MOLECULAR APPROACH TO DETERMINING THE OCCURRENCE OF PATHOGEN INDICATORS IN COMPOST

SUNAR, N. M.*[+], STEWART, D.I.*, STENTIFORD, E.I.*, AND FLETCHER, L.A*

*Pathogen Control Engineering (PaCE) Institute, School of Civil Engineering, University of Leeds, Leeds, LS2 9JT, United Kingdom.
[+]Corresponding Author. Tel: +44 (0) 113 3432319.
Email:shuhaila@uthm.edu.my

SUMMARY: An accurate method for enumerating pathogen indicators, such as Escherichia coli (*E. coli*), and *Salmonella* spp. is important for assessing the safety of compost samples. This study aimed to determine the occurrence of pathogen indicators in compost samples by using a molecular approach known as Polymerase Chain Reaction (PCR). The DNA sample was extracted from sewage sludge compost. The specificity of the probes and primers at the species level were verified by performing NCBI-BLAST2 (Basic Local Alignment Search Tool). Primers that target the *gadAB* gene for *E.coli* and *invA* gene for *Salmonella* spp. were selected which produce fragment lengths around 670bp and 285bp respectively. The primers were tested against bacterial cultures of both species and produced a strong signal band of the expected fragment length. It provided results within 6 hours which is relatively rapid compared to conventional culturing techniques. The other advantages of PCR are shown to be its high sensitivity, and high specificity.

## 1. INTRODUCTION

Composting is a technique that could help reduce a range of environmental problems associated with the treatment of biodegradable wastes. It is a good technique for preventing or reducing as far as possible the negative effects of landfill by either diverting the waste for use elsewhere or by reducing its potential for producing methane within the landfill itself. (Larney, 2003, DEFRA, 2007). England has increased its recycling rate from 7% to 27% over last 8 years but despite this improvement, more than 62% of all municipal solid waste (MSW) generated in England is disposed of in landfills (DEFRA, 2007). Waste is a resource from which compostable materials can be extracted and energy-rich fuels can be produced, reducing the amount that requires disposal to landfill. According to the Landfill Directive, European Union countries should aim to decrease the quantity of organic waste going to landfill sites to 75% of the 1995 figure by 2010, 50% by 2013 and 35% by 2020 (DEFRA, 2008). This directive could result in a large increase in the use of composting for organic wastes.

The concern over pathogens in compost has been increasing over time (de Bertoldi et al., 1983, Diaz, 2007). The presence of commonly used indicators, *E. coli* and *Salmonella* spp, has been reported in municipal solid waste (Hassen et al., 2001, Wéry et al., 2008, Ivone Vaz-Moreira, 2007, Déportes, 1998), sewage sludge (Pereira-Neto et al., 1986, D J Dudley, 1980, Bustamante et al., 2008b) and animal waste (Hanajima et al., 2006, Imbeah, 1998, Turner, 2002, Larney, 2003). When composting is managed properly, it should be able to significantly reduce the pathogen level (Larney, 2003, Hassen et al., 2001, Heinonen-Tanski et al., 2006).

According to the UK composting standard, PAS 100 (BSI, 2005) composts are required to be free of *Salmonella* spp. and to contain fewer than 1000 colony forming units of *E. coli* per gram of material. The standard cultivation methods take approximately 2-3 days to prove the existence, or otherwise, of pathogens in compost (Sidhu et al., 2001, Déportes, 1998). These 'conventional' microbiological methods have several disadvantages. They can be time-consuming, are not always specific, only report viable organisms which can be cultured (viable but non-culturable organisms – VBNC – are not reported), and sometimes fail to detect *Salmonella* spp. when present (Novinscak et al., 2007, Lee et al., 2006, Kato and Miura, 2008).

Faced with these limitations the potential of using a different approach involving a culture independent molecular-based method, called polymerase chain reaction (PCR), was investigated during this study. PCR is a technique in molecular biology, developed by Kary Mullis in 1983 (McPherson, 2000), that is highly sensitive and target specific, and more advanced versions (comparative PCR and quantitative PCR) can quantify the amount of target in a sample (M. Lebuhn, 2004). PCR allows the physical separation of any particular DNA sequence of interest then provides *in-vitro* amplification of this sequence virtually without limit (Mullis, 1990, Mullis et al., 1987). In PCR, the standard reaction uses two oligonucleotide primers which are complimentary to opposite DNA strands and flank the region of interest in the target DNA (Eeles et al., 1992). The typical range of primers used are 20-24 nucleotide-long as this length is selective enough to specify a single site in a genome of high complexity (Hayashi, 1994).

It should be possible to apply this molecular technique to the composting process in order to give more accurate results about pathogen kill in compost produced using different process conditions. The aim of this work was to evaluate the original method used in the composting quality standard method but produce results for *Salmonella* spp. and *E. coli* more rapidly using PCR.

## 2.0 MATERIALS AND METHODS

### 2.1 Raw material

Primary sewage sludge was collected from the main city wastewater treatment plant in Leeds, UK and mixed with matured compost produced from kitchen waste. The composting mixture was prepared by adding sewage sludge and matured compost from kitchen waste in the ratio 6:4 (w/w). The compost mixture was shredded to an average size of 5 – 20 mm (Hu et al., 2009) before it was placed in a laboratory compost reactor.

### 2.2 Laboratory-scale composting

The laboratory composting apparatus used was the same as that used in the DR4 biodegradability tests method devised by WRc (Waste Research Centre) in the United Kingdom. The material (compost mixture) was incubated under aerobic conditions in a reactor using forced aeration. The composting reactor was cylindrical (22 cm height x 8 cm diameter) with a perforated plate at the bottom to distribute the air supplied, and was loaded with 400 g of the mixture. The air was supplied using a pump at a constant flow rate of 0.5L/min, which was measured and controlled using a flow meter (tube-and-float type). In this sealed reactor, the air was forced up through the compost material, and passed out of the reactor. Then the air was passed through a condenser, to remove surplus liquid before exhausting to atmosphere. The temperature of the compost material was monitored with a thermocouple which was inserted in the sample.

## 2.3 Conventional enumeration approach

The conventional method for the enumeration of *Salmonella* spp. and *E. coli* was carried out using serial dilution followed by a standard membrane filtration technique. This method is recommended by the UK compost quality standard method PAS 100 (BSI, 2005). The enumeration of Salmonella spp. (BS EN ISO 6579:2002) used the Muller-Kauffmann tetrathionate/novobiocin broth as a resuscitation medium followed by growth on Rambach agar. For *E. coli*, tryptone-bile-glucuronide medium (TBX media) was used (BS ISO16649-2:2001) which is a solid medium containing a chromogenic ingredient used for detection of the enzyme –glucuronidase in *E. coli* cells. Compost sample preparation involved taking a 25 g sub-sample and placing this into a stomacher bag together with 225ml of sterile PBS (phosphate buffered saline). This sample was then stomached for 60 seconds and subject to serial 10-fold dilution of the resulting supernatant using PBS.

## 2.4 Molecular biology approach

### 2.4.1 Polymerase chain reaction procedure

The original PCR experiments were carried out on pure cell cultures to validate the functionality of the reagents and primers. The cells were taken from an individual colony on the surface of an agar plate using a sterile toothpick and resuspended in 100μl of sterile distilled water. Samples were heated at 99 °C for 5 minutes and centrifuged at a top speed in a micro centrifuge for 1 minute to remove cell debris. The supernatant was then transferred into a new tube to be used as a source of DNA. The procedure for collecting *E.coli* and *Salmonella* spp. DNA from the compost mixture was the same as outlined above with a sterile toothpick being used to take a sample from the suspension produced by stomaching. The compost suspension sample was also resuspended in 100μl and subjected to the same heating and centrifuging methods outline above.

The PCR reaction mixture contained 2.5μl of DNA solution from the procedure above, 5 units of GoTaq reaction buffer (from Promega Corp., USA), 1 x PCR reaction buffer, 1.5mM $MgCl_2$ (already in the GoTag reaction buffer), 10 mM PCR nucleotide mix (Promega Corp., USA), and 1.5 μM DNA primer in a final volume of 50 μl. The reaction mixtures were incubated at 95°C for 2 min, and then cycled 30 times for another three steps: denaturing (95°C, 30s), annealing (50°C, 30s), primer extension (72°C, 45s). This was followed by final extension step at 72°C for 7 min. Amplification product sizes were verified by electrophoresis of 10 μl samples in a 1.0% agarose TBE gel with ethidium bromide straining (Stewart et al., 2007).

### 2.4.2 16S rRNA Gene Sequencing: Primer Selection

The desired PCR products were approximately 670 bp for *E. coli* and 285 bp for *Salmonella* spp. in length. Those two sets of PCR primers were selected from previous studies (D. De Clercq, 2007, McDaniels, 1996, Rahn et al., 1992) as described in Table 1. The specificity of all probes and primers at the species level were verified by performing the NCBI-BLAST2 (Basic Local Alignment Search Tool) by EMBL Nucleotide Sequences Database (European Molecular Biology Laboratory) from the European Bioinformatics Institute. Default settings were used for the BLAST parameters (match/mismatch scores 2,-3, open gap penalty 5, gap extension penalty 2).

Table 1: DNA sequences used for the PCR primer and probes (Rahn et al., 1992, De Clercq, 2007, McDaniels, 1996)

| Primer and probes | Sequences (5'→3') | Microorganisms | Genes | Functions | GenBank accession no. |
|---|---|---|---|---|---|
| *E. coli* forward *gadA/B* primer | ACC TGC GTT GCG TAA ATA | *Escherichia coli* | *gad A/B* | Glutamate decarboxylase | M84024 (*gadA*) M84025 (*gadB*) |
| *E. coli* reverse *gadA/B* primer | GGG CGG GAG AAG TTG ATG | *Escherichia coli* | *gad A/B* | Glutamate decarboxylase | M84024 (*gadA*) M84025 (*gadB*) |
| invA139 | GTG AAA TTA TCG CCA CGT TCG GGC AA | *Salmonella* spp. | *invA* | Invasion protein | M90846 |
| invA141 | TCA TCG CAC CGT CAA AGG AAC C | *Salmonella* spp. | *invA* | Invasion protein | M90846 |

## 2.5 Analysis of compost characteristics

A number of different characteristics of the compost samples were also determined. The pH values were determined from a compost/water suspension extract according to the procedures described by Hu et al., (2009). The moisture content of the compost material was determined by drying a 20g sample at 105°C for 24h hours. The drying, cooling and weighting process were repeated until no change in weight was observed.

## 3.0 RESULTS AND DISCUSSION

### 3.1 Characteristics of the raw material

Table 2 shows the characteristics of the sewage sludge and kitchen waste derived compost material used to produce the compost mixture used in this investigation. The sewage sludge and kitchen waste derived compost material had a moisture contact of 67% and 52% respectively. According to Haug (1980), the optimum moisture content recommended values are in the 50% to 70%. The moisture content of the waste mixture was maintained in the 50% to 70% range during the whole of the composting process using distilled water lightly sprinkled onto the sample as required. The moisture content is important to provide a medium for the transport of dissolved nutrients required for the metabolic and physiological activities of microorganisms (Miller, 1989, Stentiford, 1996).

## 3.2 Characteristics of composts

Figure 1 shows the temperature profiles of the laboratory-scale composting test. The temperature was increased from ambient to the thermophilic range over a 2 day period. It was then kept at that temperature until day 8. According to Deportes et al., (1995) temperatures of approximately 55-60 °C for at least 3 days are recommended to be effective against *Salmonella* spp. and *E. coli*. The temperature profiles in the laboratory scale composters were set to at least exceed this level of exposure as can be seen from Figure 1. The pH values were around neutral at the beginning of the composting process, and then increased as shown in Figure 1 reaching 8.8 at the end of the composting trial. This increase of pH is characteristic of many composting operations due to ammonia produce from protein decomposition. (de Bertoldi et al., 1988, Haug, 1980). The moisture content in the composting process is affected by the heat generated in the mass and the amount of aeration used, these also affect the temperature and the rate of decomposition (Turner, 2002). In this study, it was found that the moisture content of the final compost was in the range 40-50% which was slightly lower than that observed in the initial mixtures.

Table 2: Characteristic of sewage sludge, kitchen waste compost and their mixture [a]

| Sample | pH | Moisture Content (%) |
|---|---|---|
| Sewage Sludge | 5.7 | 67 |
| Kitchen waste derived compost | 6.6 | 52 |
| Combined mix (Sewage sludge : Kitchen waste, 6:4 w/w, wet weight) | 6.2 | 60 |

[a] Data are mean values of three replicates

## 3.3 The occurrence of *Salmonella* spp. and *E. coli*

Figure 2 shows the change in the numbers of *Salmonella* spp. and *E. coli* when enumerated using the conventional method (membrane filtration). The laboratory scale study showed that after composting for 2 days (which included at least 1 day at 55°C) the numbers for both species were below the limits in the UK compost standard (PAS 100) which requires the compost to be free of *Salmonella* spp. and to contain fewer than 1000 colony forming units of *E. coli* per gram of material. Both targets were reached within 2 days, however, the conventional enumeration methods typically took between 2 to 3 days to complete. However PCR analysis undertaken on the same samples yielded results much faster, in this case within 6 hours.

Figure 3 shows the PCR results for 4 samples of compost material. The gel image shows that *E. coli* (DNA template) formed a 670bp PCR product with the *E .coli gad AB* gene primer pair (lines 3 to 6), and *Salmonella* spp. (DNA template) formed a 285bp PCR product with the *Salmonella* spp. i*nv A* gene primer pair (line 8). Thus *E. coli* was detected in all samples during the composting process (positive bands in Lines 3 to 6) whereas *Salmonella* spp. were detected only on the initial day of composting (positive band shown in line 8) and not in subsequent samples on day 2, day 5 and day 8 (lines 9 to 11). These results were confirmed by the membrane filtration culture method. Table 3, shows the numerical results given by the membrane filtration method, but, it took much longer to obtain results than using PCR. The results using PCR were obtained within 6 hours but were only showed the presence or absence of the target organisms and did not give the numbers present.

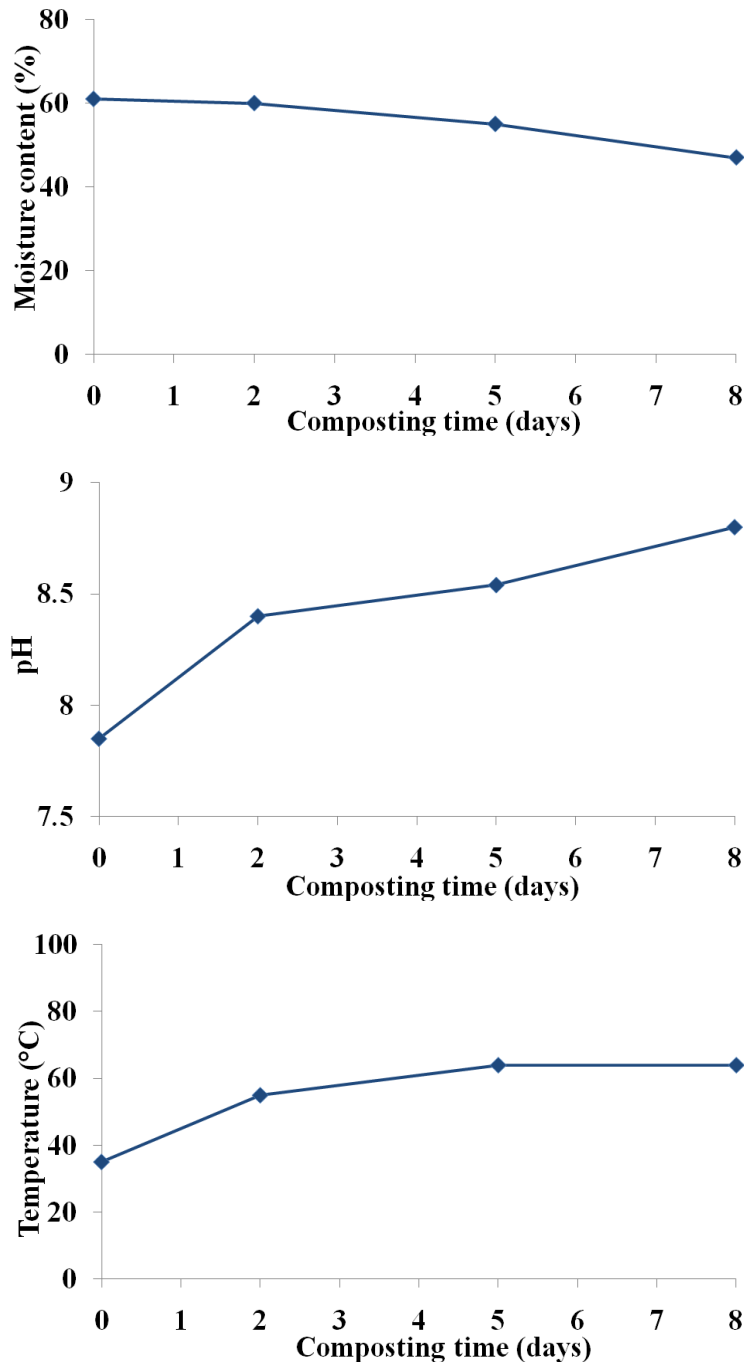

Fig. 1: Moisture content, pH changes and temperature profiles of the laboratory-scale of composting test

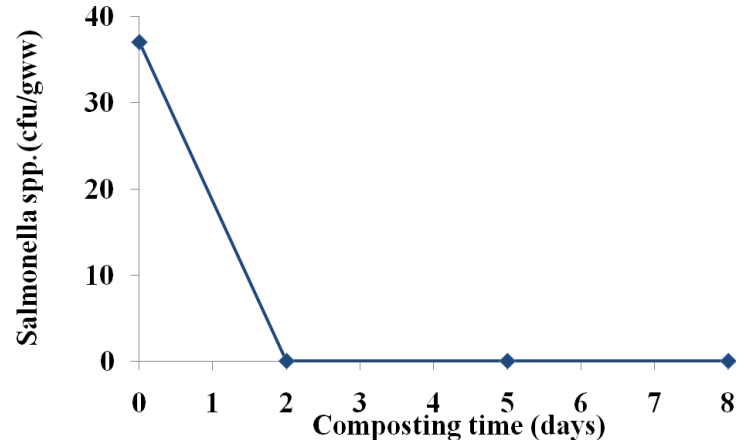

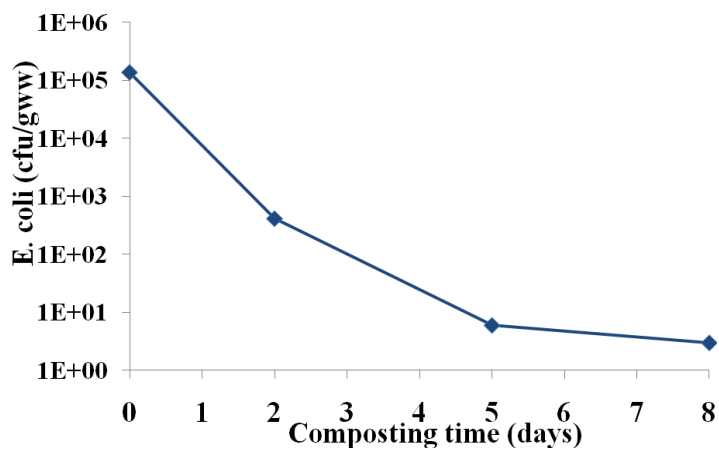

*Figure 2: Enumeration of Salmonella spp. and E. coli by using conventional method*

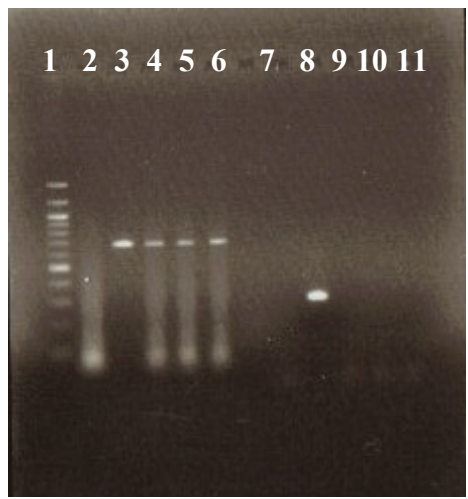

*Figure 3: Line 1: Ladder (100bp New England Bio); Line 2: E. coli control; Line 3: E. coli DNA template; 35°C with 0 day of composting; Line 4: E. coli: 55°C, day 2 ; Line 5: E. coli: 64°C , day 5; Line 6: E. coli: 64°C, day 8; Line 7: Salmonella spp. control; Line 8: Salmonella spp.: 35°C, 0 day; Line 9: Salmonella spp.: 55°C, day 2 ; Line 10: Salmonella spp.: 64°C , 5 day ; Line 11: Salmonella spp.: 64°C, day 8.*

Table 3: Summarizes results for both methods in determines the occurrences of pathogens.

| Conditions of composting | Membrane filtration [a] | | PCR [b] | |
| --- | --- | --- | --- | --- |
| | E. coli (cfu / gww) | Salmonella spp. (cfu / gww) | E. coli | Salmonella spp. |
| 35°C, 0 days | 136000 | 37 | Yes | Yes |
| 55°C, 2 days | 412 | 0 | Yes | No |
| 64°C, 5 days | 6 | 0 | Yes | No |
| 64°C, 8 days | 3 | 0 | Yes | No |

[a] Data are means values of three replicates; membrane filtration procedure needs 2-3 days to complete.
[b] PCR procedure needs approximately 6 hours to complete.

## 4.0  CONCLUSIONS AND RECOMMENDATIONS

The results show that the PCR procedure can be used as a rapid molecular approach to determine the occurrence of pathogen indicators in compost material. The occurrence of *inv A* for *Salmonella* spp. and *gad AB* gene for *E. coli* was successfully detected by PCR and confirmed by conventional techniques (membrane filtration). The study also showed that this PCR method gives results much faster than the conventional method, typically within 6 hours. However, the results from the PCR only showed the presence or absence of the target microorganisms and not their concentrations.

Further study will give a better understanding of the degree of specificity of DNA primers in compost materials. Work is currently in progress which will use a variation of this PCR method to give quantitative results for the target microorganisms.

ACKNOWLEDGEMENT

The authors wish to thank Dr. Pete Hobbis and the Public Health Laboratory staff, School of Civil Engineering, University of Leeds for their valuable support and excellent laboratory facilites. This study was funded by Ministry of Higher Education of Malaysia and University of Tun Hussien Onn Malaysia